\documentclass[aps,pre, reprint,superscriptaddress, showkeys]{revtex4-1}
\usepackage{amsmath}
\usepackage{xcolor}
\usepackage{graphicx}
\usepackage{ulem}
\usepackage{amsfonts}				
\usepackage{float}					
\usepackage{longtable}
\usepackage{array}
\usepackage{hyperref}

\newcommand{\norm}[1]{\left\vert#1\right\vert}
\newcommand{\mb}[1]{\mathbf{#1}}

\begin{document}
	
\title{Predictability in Rotating Turbulence: Insights from a Shell Model Study}
\author{Shailendra K. Rathor}
\email{skrathor@iitk.ac.in}
\affiliation{Department of Physics, Indian Institute of Technology Kanpur, Uttar Pradesh 208016, India}
\begin{abstract}
	We investigate the predictability aspects of rotating turbulent flows through extensive numerical simulations of a shell model of rotating turbulence. In particular, we measure the large-scale predictability time and find that it increases with rotation rate to satisfy a power law in Rossby number with a scaling exponent of $-2/3$. Intriguingly, we find that before entering the algebraic growth stage, the error dynamics freezes for a time period determined by the finite Rossby number. We further analyse the scale dependence of the predictability time and observe that it tends to become scale independent in the Zeman range as the Rossby number decreases. Finally, we compute the finite size Lyapunov exponent and validate the dimensional prediction of its scaling $\sim\delta^{-1}$ for large $\delta$ of the order of velocities in the Zeman range for small Rossby numbers.
\end{abstract}
\maketitle
	
\section{Introduction}
Predictability in turbulence~\cite{Bohr1998,Cencini2009} is an important aspect of certain practical problems, such as weather forecasting~\cite{Lorenz1969,Leith1971,Leith1972}. This is because turbulence does not lose its predictability due to its inherent chaotic multiscale dynamics, even though it is regarded as a highly chaotic phenomenon.
This enables one to make long-range weather predictions. Although the predictability in three-dimensional (3D) and two-dimensional (2D) turbulence has been extensively studied~\cite{Boffetta1997,Boffetta2001,Boffetta2017,Berera2018}, the study of predictability in rotating turbulence~\cite{Godeferd2015,Greenspan1968}, which is observed in many natural systems~\cite{Barnes2001,Cho2008,Aurnou2015} and is a key factor in weather prediction models~\cite{Bartello1995,Vannitsem2017}, is relatively limited\cite{Ngan2009} and thus requires further investigation.

{The predictability of the future state of a chaotic system with complete knowledge of the evolution laws is severely limited by the sensitive dependence on initial conditions and hence by an imperfect knowledge of the present state. For example, the predictability in turbulence is limited by the error in the state caused by the velocity fluctuations---induced by the thermal fluctuations representing the chaotic molecular motions in the fluid---at the fastest timescales near the dissipation scale~\cite{Ruelle1979,Crisanti1993a}.} 
{In the phase space of the chaotic system, the distance $\delta(t)$ between two trajectories initially separated by an infinitesimal error $\delta_0$  increases exponentially as $\delta(t) \simeq \delta_0 \exp(\lambda t)$ with a typical growth rate $\lambda$ known as the maximum Lyapunov exponent (in the limit of $t \rightarrow \infty$). Therefore, the future state of the chaotic system is expected to be predictable within a tolerance error $\delta_{\rm max}$ of interest up to a time, the predictability time $T$, of the order of $\lambda^{-1}$:
\begin{equation*}
T \sim \frac{1}{\lambda} \ln \norm{\frac{\delta_{\rm max}}{\delta_0}}.
\end{equation*}
}

In many situations, such as turbulence, the predictability time based on the maximum Lyapunov exponent becomes irrelevant and insignificant as compared to the observed predictability time~\cite{Bohr1998}.

Turbulence has many degrees of freedom with different characteristic lengthscales and timescales (eddy turnover times)~\cite{Aurell1996,Lorenz1969}, which interact nonlinearly with each other: the infinitesimally small error at the Kolmogorov scale grows exponentially and propagates to the larger scales due to nonlinear interactions. Thus, the error at relatively large scales does not remain infinitesimally small, and hence the predictability time determined by the maximum Lyapunov exponent is insufficient and irrelevant for predicting the state of a turbulent system at large scales of interest~\cite{Aurell1996a}.

Lorenz~\cite{Lorenz1969} proposed, based on his physical arguments stemmed from the assumptions in the energy cascade picture of turbulence, that the time $\tau(k)$ it takes a non-infinitesimal error at scale $l_0 \sim 1/2k$ to cause complete uncertainty (by growing and transferring) at the larger scale $l \sim 1/k$ is proportional to the characteristic eddy turnover time at scale $l$. In simpler terms, the evolution of non-infinitesimal error at different scales is governed by the corresponding characteristic timescales~\cite{Lorenz1969,Leith1972,Ruelle1979}. 
A small perturbation (error) initially at Kolmogorov scale $\eta \sim 1/k_d$, where $k_d$ is dissipation wavenumber, grows and propagates to the scale of interest $L \sim 1/k_L$ of large energy-containing eddies through the process of an inverse cascade, and hence the predictability time $T_L$ at large scale $L$ is the sum of all the characteristic times required for the small perturbation to propagate from Kolmogorov scale $\eta$ to the large scale of interest $L$: $T_L = \sum_{p=0}^{P} \tau(2^{p} k_L )$, where $P = \log_2 (k_d/k_L)$. 
For the turbulent flows with high Reynolds number and spectral velocity $\norm{u(k)} \sim k^{-\alpha}$, the eddy turnover timescale $ \tau(k) = \left[k \norm{u(k)}\right]^{-1} \sim k^{\alpha -1}$ and hence the predictability time $T_L$ is estimated~\cite{Bohr1998,Ditlevsen2010} as

\begin{eqnarray}
T_L \sim \sum_{p=0}^{\infty} \tau(2^{p} k_L ) = k_L^{\alpha -1} \frac{1}{1-2^{\alpha -1}}.
\label{eqn:Lorenz_approach}
\end{eqnarray}

In case of 3D fully developed turbulence, $\alpha = 1/3$ because
$\norm{u(k)} \sim k^{-1/3}$ and so $T_L \sim k_L^{-2/3} \sim \tau(k_L)$.
This immediately implies that the major contribution of predictability time in Eq.~(\ref{eqn:Lorenz_approach}) comes from the term corresponding to $p = 0$. 
Therefore, the predictability time in 3D turbulence, from the algebraic expression in Eq.~(\ref{eqn:Lorenz_approach}), is determined by the eddy turnover time of the largest scale of interest, and the predictability time corresponding to the maximum Lyapunov exponent becomes irrelevant. 
The more energy there is at small scales, faster the error propagates to the large scales, resulting in a shorter predictability time. 
In 2D turbulence, $\alpha = 1$ because the energy spectrum in the small scales $E(k) \sim k^{-3}$ and $\norm{u(k)}^2 = \int E(k) {d}k$; this results in diverging predictability time $T_L$. 
As a consequence, long-range predictability is realizable in large-scale atmospheric flow that is modeled as a 2D turbulent system.

The predictability behaviour of rotating turbulence is expected to be different from 3D homogeneous isotropic and 2D turbulence because rotation introduces a new timescale and a characteristic lengthscale called the Zeman scale~\cite{Zeman1994} in the problem, and new phenomena emerge in presence of rotation~\cite{Smith1999,Bartello1994,Sreenivasan2008}.
The energy spectrum $E(k) \sim k^{-2}$ for the scales in the Zeman range $(k < k_\Omega)$, where $k_\Omega$ is the Zeman wavenumber corresponding to the Zeman scale, and the energy spectrum $E(k) \sim k^{-5/3}$ for small scales $(k > k_\Omega)$~\cite{Mininni2012,Biferale2016,Rathor2020}. 
The predictability in rotating turbulence is unlikely to change significantly considering it is limited by the small-scale dynamics as in 3D turbulence. 
However, because $\alpha = 1/2$ in large scales, the predictability time is expected to modify to $T_L \sim k_L^{-1/2} \sim \tau_{rot}(k_L)$, where $k_L < k_\Omega$ and $\tau_{rot}(k) \sim k^{-1/2}$ is the eddy turnover time in the scales ($k < k_\Omega$).
Since $\tau_{rot}/\tau \sim k^{1/6}$, this immediately implies that the predictability time at large scales in the Zeman range $k < k_\Omega$ in rotating turbulence is \textit{larger} than that in 3D homogeneous isotropic turbulence for the same scales. 

In this work, we investigate the behaviour of large-scale predictability, using our extensive shell model simulations, by measuring the growth of the error between two initially close trajectories in the state space of the shell model, and thus test the aforementioned predictions about predictability in rotating turbulence.
First, we look at the large-scale predictability of the full system that takes all components (scales) into consideration, and how it relates to the Rossby number, a non-dimensional parameter that characterises the  rotating turbulence.
We then look into the scale dependence of predictability for different Rossby numbers. Finally, we compute the finite size Lyapunov exponent as a measure of predictability and validate the dimensional predictions. 


\section{Simulation Details}
We consider a turbulent flow, under a solid body rotation $\mb{\Omega}$, described by the Navier--Stokes equation for the velocity field $\mb{u}$ of a 3D incompressible ($\mb{\nabla}\cdot \mb{u} = 0$) flow
\begin{equation}
\frac{\partial \mb{u}}{\partial t} + (\mb{u}.\mb{\nabla}) \mb{u}  =   -\frac{1}{\rho} \mb{\nabla} p + \nu \nabla^2 \mb{u} - 2 \mb{\Omega} \times \mb{u} + \mb{f},
\label{eqn:NSE}
\end{equation}
where $\rho$ is the fluid density and $\nu$ denotes the kinematic viscosity of the fluid. The centrifugal force, which is absorbed in the pressure field $p$, and the Coriolis force $-2 \mb{\Omega} \times \mb{u}$ are generated by the solid body rotation. Furthermore, turbulence in the flow is sustained by the external force $\mb{f}$, which pumps constant energy $\langle \mb{u}\cdot \mb{f} \rangle$ at large scales into the system.

To study the predictability aspects of the aforementioned rotating turbulent flow, we numerically simulate shell models of turbulence modified appropriately to take the Coriolis force into account.
Shell models are innately low-dimensional dynamical systems that resemble the spectral Navier--Stokes equation but actually are not derived from
it~\cite{Frisch1995,Bohr1998,Biferale2003,Pandit2009}. The dynamical system is constructed by
dividing the 3D Fourier space into discrete non-overlapping shells such that the wave vectors in the shell $n$ are represented by a single wavenumber $k_n$ that live on a one-dimensional logarithmically-spaced lattice: $k_n = k_0 \lambda^n$, where $k_0$ and the shell spacing $\lambda$ are real constants. A complex dynamical variable $u_n$, which represents the velocity difference associated with a lengthscale $\sim k_n^{-1}$ in the Navier--Stokes equation, is associated with the shell $n$ in the lattice. The nonlinear interactions that satisfy the important symmetries of the Navier-Stokes equations are chosen and  restricted to the nearest and next-nearest neighbour shells. Because of this logarithmic construction on a one-dimensional lattice truncated with a total of $N$ shells and restricted nonlinear interactions, extremely high Reynolds numbers that are not possible in DNSs are achieved in shell models even with a few shells, i.e., small $N$.

Shell models have extensively been used to discover and reproduce many characteristics of predictability in turbulence~\cite{Crisanti1993,Crisanti1993a,Crisanti1994,Aurell1996,Aurell1996a,Bohr1998}. However, the use of shell models in the study of rotating turbulence is scarce and relatively recent~\cite{Hattori2004,Chakraborty2010,Rathor2020,Rathor2021}. Furthermore, shell models that are structurally isotropic are able to reproduce and predict many features of rotating turbulence, such as the
dual scaling of energy spectrum, two-dimensionalisation, and the scaling of equal-time structure
functions~\cite{Hattori2004,Chakraborty2010,Rathor2020}. 

We, therefore, resort to shell models to study predictability in rotating turbulence and consider the following so-called GOY shell model~\cite{Gledzer1973,Ohkitani1982} modified for rotating turbulence
\begin{eqnarray}\label{goy}
\frac{du_n }{dt}= \mathcal{N}_n -\nu k_n^2 u_n  +f_n -i\Omega u_n,
\end{eqnarray}
where  $\mathcal{N}_n$ is the nonlinear term modeled as
\begin{eqnarray}
\mathcal{N}_n& = & i \left[a k_{n+1} u_{n+2} u_{n+1} + b k_{n} u_{n+1}u_{n-1} \right.\nonumber \\
&& \left.+ c k_{n-1} u_{n-1} u_{n-2}\right]^{*} \nonumber
\end{eqnarray}
with real coefficients $a, b$ and $c$, and the superscript $^*$ denoting a complex conjugate. {We set the coefficient $a = 1$ for time rescaling without any loss of generality, and we determine the other interaction coefficients as $b = -1/2$ and $-1/2$ from the conservation of quadratic invariant quantities, namely energy and helicity, for inviscid and unforced 3D turbulence.} In Eq.(\ref{goy}), the second term on the right is the viscous dissipative term with viscosity denoted by $\nu$, and 
$f_n$ is the forcing term to sustain the turbulence. We take the following forcing scheme to supply constant energy and zero helicity in the system~\cite{Ditlevsen2001}:
$f_n = \epsilon (1+i) \left(\frac{u_n}{\norm{u_n}^2}\delta_{n,2} + \frac{u_n}{2\norm{u_n}^2}\delta_{n,3}\right)$,
where $\epsilon$ determines the energy injection rate. {The total mean energy injection rate is $\bar{\epsilon} = 1.5 \epsilon$ and, in our simulations, $\epsilon = 0.01$.}
Finally, the last term with rotation rate $\Omega$ acts as Coriolis force. This term is specifically made imaginary so that it does not explicitly contribute to the kinetic energy of the system.

We simulate the aforementioned GOY model of rotating turbulence with a total of $N = 30$ shells, $\lambda = 2$ shell spacing, and $k_0 = 1/16$. A high Reynolds number $Re\sim 10^{10}$ is achieved by choosing the viscosity $\nu = 10^{-9}$ such that the Kolmogorov shell $n_\eta = 24$. Because the shell model is a set of stiff ordinary differential equations, the integration is performed carefully using the RK4 scheme with a time step as small as $\delta t = 10^{-5}$. We take the initial velocity field $u_n = \sqrt{k_n} \exp(i\theta)$ for $n\le 4$
and $u_n = \sqrt{k_n} \exp(-k_n^2) \exp(i\theta)$  for $n \ge 5$, where $\theta \in [0,2\pi]$ is a random phase. We evolve the system to bring it to a statistically steady state before starting the simulations for predictability. 

We simultaneously and independently evolve two initially close trajectories of reference velocity $u_n(t)$ and perturbed velocity $u_n^\prime(t)$ in the $2N-$dimensional state space of the shell model. The initial perturbed velocity $u_n^\prime(t = 0)$ is obtained by perturbing initial reference velocity $u_n(t = 0)$ that is in a steady state, in accordance with the following scheme\cite{Aurell1996}:

\begin{equation}
u_n^\prime = 
\begin{cases}
\exp(i\phi_n) u_n, & n \geq n_{\eta}\\
u_n, & n < n_{\eta},
\end{cases} 
\end{equation}
where $n_{\eta}$ is the shell number corresponding to the Kolmogorov wavenumber $k_\eta$, $\phi_n \in [0, \theta_c] $ is a random number, and $\theta_c$ quantifies correlation between the two fields. Initially, the energy contents  $\frac{1}{2}\sum_{n=1}^{N} \norm{u_n}^2$ and $\frac{1}{2}\sum_{n=1}^{N} \norm{u_n^\prime}^2$ are equal, however, the error measured as the (Euclidean) distance between $u_n$ and $u_n^\prime$ is non-zero.

To study the predictability, we track the evolution of the error field 
$\delta u_n(t) = \left(u_n^\prime(t) -u_n(t)\right)/\sqrt{2}$,
where the factor $1/\sqrt{2}$ is for normalization convenience. The \textit{error} between the two trajectories is measured from $\delta u_n(t)$ in terms of the energy norm, defined as
\begin{equation}
E_\Delta(t) = \int_{0}^{\infty} E_\Delta(k,t) \mathrm{d}k = \frac{1}{2} \sum_{n=1}^{N} \norm{\delta u_n(t)}^2,
\end{equation}
where $E_\Delta(k,t)$ is the error spectrum. In shell model, the error spectrum can be defined as
\begin{equation}
E_\Delta(k_n,t) = \frac{\norm{\delta u_n(t)}^2}{k_n}.
\end{equation}

The initially correlated trajectories $u_n(t)$ and $u_n^\prime(t)$ evolve to completely decorrelate in time, such that the error $E_\Delta(t)$ grows to eventually saturate to $E(t)$, where $ E(t) = \sum_{n=1}^{N} \norm{u_n(t)}^2 = \sum_{n=1}^{N} \norm{u_n^\prime(t)}^2$. In our simulations, we choose fairly small $\theta_c (= 10^{-6})$ to generate initially strongly correlated trajectories $u_n(t)$ and $u_n^\prime(t)$.

We repeat our simulations with various rotation rates, turning on the Coriolis force term simultaneously for both the reference and the perturbed trajectories.
A flow in a rotating frame is characterised by a non-dimensional Rossby number in addition to the Reynolds number. The Rossby number corresponding to rotation rate $\Omega$ is defined as $\mathrm{Ro}:=U/\Omega L_{0}$, where $U= (\sum_{n} \mid u_n \mid ^ 2)^{1/2}$ is the root-mean-square velocity and $L_{0}= 1/k_0$. For finite $\mathrm{Ro}$, we determine $U$ at late times when the system is reached in (asymptotically) steady state. We report our results for moderate Rossby numbers: $\mathrm{Ro} = 0.137, 0.049, 0.025$, and $0.018$ corresponding to the rotation rates $\Omega = 1, 4, 10$, and $16$ respectively, in addition to $\mathrm{Ro} =  \infty$ (i.e., no rotation). The statistics, for each $\mathrm{Ro}$, is performed over an ensemble of $5000$ different realizations of the pairs of reference and perturbed trajectories in the state space of the shell model. 

The characteristic timescale of the system, the large eddy turnover time $T = L_0/U$, varies in the simulations with different Ro because the timescales in the Zeman range are governed by the Coriolis force. However, the timescale of the dynamics at the Kolmogorov scale is not significantly affected for moderate Ro. Therefore, we consider the Kolmogorov timescale $t_\eta = \sqrt{\nu/\varepsilon}$, where $\varepsilon$ ($= \bar{\epsilon}$, in steady state) is the energy dissipation rate, as the characteristic timescale for all of the simulations and report our results in units of the constructed timescale $T_{\eta} = 10^4 t_\eta$. In our simulations, for $\mathrm{Ro} = \infty$, $T \simeq 6T_{\eta}$.


\section{Results}
\subsection{Error Dynamics}
\begin{figure}
	\includegraphics[scale=0.98]{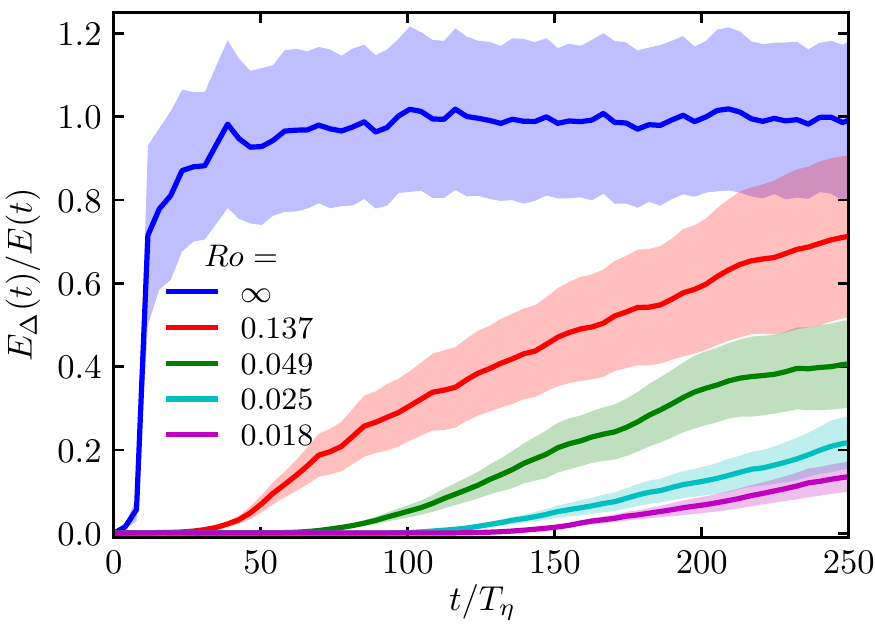}
	\caption[Plots of the relative error growth with time.]{Plots of the relative error $E_\Delta(t)/E(t)$ versus time $t$ normalized with $T_{\eta} = 10^4 t_\eta$, where $t_\eta$ is the Kolmogorov time, for various $\mathrm{Ro}$ values (see legend). Solid lines with colors corresponding to the respective $\mathrm{Ro}$ represent the mean relative error, whereas the respective shaded region represents the fluctuations of the relative error within half a standard deviation (for clarity to the eye) from the mean. }
	\label{fig:error_energy}
\end{figure}

In Fig.~\ref{fig:error_energy}, we plot the time evolution of error in terms of the relative error $E_\Delta(t)/E(t)$, defined as the ensemble average of error $E_\Delta(t)$ normalized with the ensemble average of energy $E(t)$ (of reference trajectories). 
The solid lines show the mean relative error, and the respective shaded region represents the fluctuations of the relative error within half a standard deviation from the mean. The mean relative error displays an initial exponential growth followed by an algebraic growth before it saturates to $ E_\Delta(t)/E(t) \simeq 1$. As the error approaches saturation, the fluctuations in the relative error increase to become enormously large. 
This results from the kinetic energy fluctuations, which vary on the same timescale as the saturated error and are related to the large-scale dynamics. The forcing mechanism at large scales could also contribute to these large-scale fluctuations.  Moreover, we observe a decrease in the fluctuations with a decreasing Ro. This is perhaps the result of the \textit{linear} Coriolis force dominated dynamics at large scales such that the nonlinear fluctuations are suppressed.

\begin{figure}
	\includegraphics[scale=0.98]{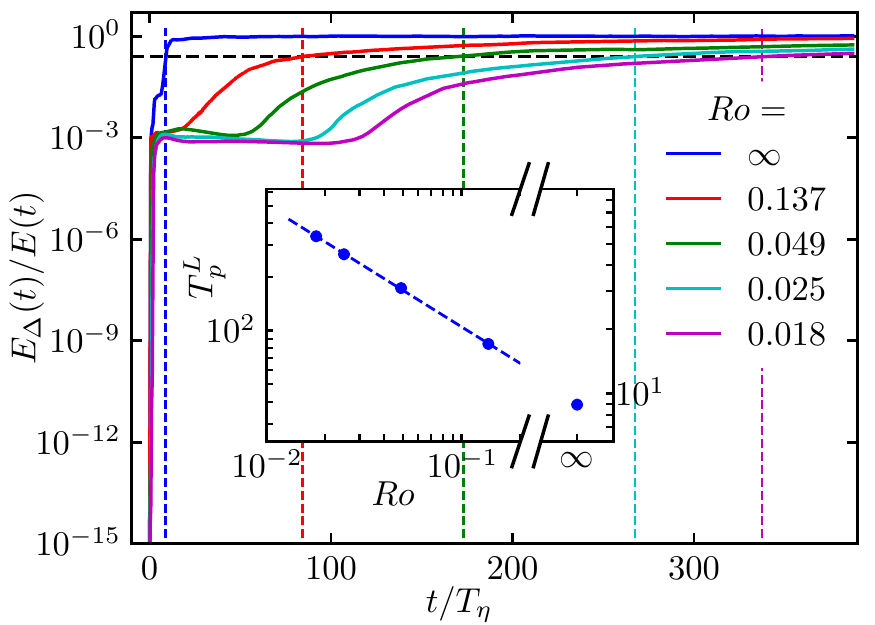}
	\caption[Semilog plots of error growth and predictability in inset figure.]{Semilog plots of the mean relative error energy $E_\Delta(t)/E(t)$ versus time $t$ normalized with $T_{\eta} = 10^4 t_\eta$, where $t_\eta$ is the Kolmogorov time, for different Rossby numbers (see legend) from our simulations of shell model. The horizontal black dashed line represents $E_\Delta(t)/E(t)=1/4$.  The vertical dashed lines with colors corresponding to the respective $\mathrm{Ro}$ determine the large-scale predictability time. Inset: Loglog plot of predictability time $T_p^L$ versus Rossby number $\mathrm{Ro}$. The blue dashed line is the linear best fit with slope $= -0.68$.}
	\label{fig:Tp_Leith}
\end{figure}

Figure~\ref{fig:Tp_Leith} presents the time evolution of the relative error on a semilogarithmic scale. For all Ro, we observe an initial stage of exponential growth of error that lasts for a relatively short time period $\sim T_{\eta}$.
For $\mathrm{Ro} = \infty$, this is followed by a short-lived linear growth stage before the error energy saturates to $ E_\Delta(t) \simeq E(t)$ in its later stage of evolution. However, peculiar to finite Ro, the error growth intriguingly freezes, after the exponential stage, for a time period that depends on the Rossby number; smaller the $\mathrm{Ro}$, longer the time period. This stage of frozen error evolution corresponds to the plateaus at the relative error $\sim 10^{-3}$ (see Fig.~\ref{fig:Tp_Leith}).
Furthermore, after this frozen stage, the relative error begins to develop again, but algebraically. This is as though the dynamics of the reference and the perturbed trajectories synchronise in the state space of the shell model for an Ro-dependent time period before the trajectories desynchronise and separate again but algebraically.

Predictability time associated with a scale is defined as the time a small error (at the Kolmogorov scale) takes to grow and propagate to the scale to induce complete uncertainty at the scale~\cite{Bohr1998}. Moreover, predictability time for the full system is defined as the time it takes for a small initial error to grow to a predetermined threshold set on the relative errors. The classical prescription due to Leith~\cite{Leith1971} for fixing the threshold is $E_\Delta(T_p^L)/E(T_p^L) = 1/4$ such that $T_p^L$ denotes the predictability time for the full system.
In Fig.~\ref{fig:Tp_Leith}, the vertical dashed lines with colors corresponding to the respective $\mathrm{Ro}$ determine the predictability time $T_p^L$ by intersecting the respective relative error curve at the threshold $E_\Delta(T_p^L)/E(T_p^L) = 1/4$. Interestingly, we observe that the time the relative error freezes for forms a considerable part of the predictability time $T_p^L$(see Fig.~\ref{fig:Tp_Leith}). For instance, for Ro $= 0.018$, the duration of frozen stage $\simeq 100 T_\eta$ is about $30\%$ of the predictability time $T_p^L \simeq 337T_\eta$.

In the inset of Fig.~\ref{fig:Tp_Leith}, we plot the predictability time $T_p^L$ against Rossby number $\mathrm{Ro}$ on a loglog scale and find that $T_p^L$ increases with decreasing Ro. Surprisingly, for finite $\mathrm{Ro}$, the dependence of the predictability time $T_p^L$ on $\mathrm{Ro}$ satisfies a power law $T_p^L \sim \mathrm{Ro}^{\beta}$ with the measured value of exponent $\beta = -0.68(\simeq -2/3)$, as shown by the linear best fit line (see blue dashed line in the inset of Fig.~\ref{fig:Tp_Leith}).

\subsection{Scale-by-scale Predictability}

\begin{figure*}[ht]
	\includegraphics[scale = 1]{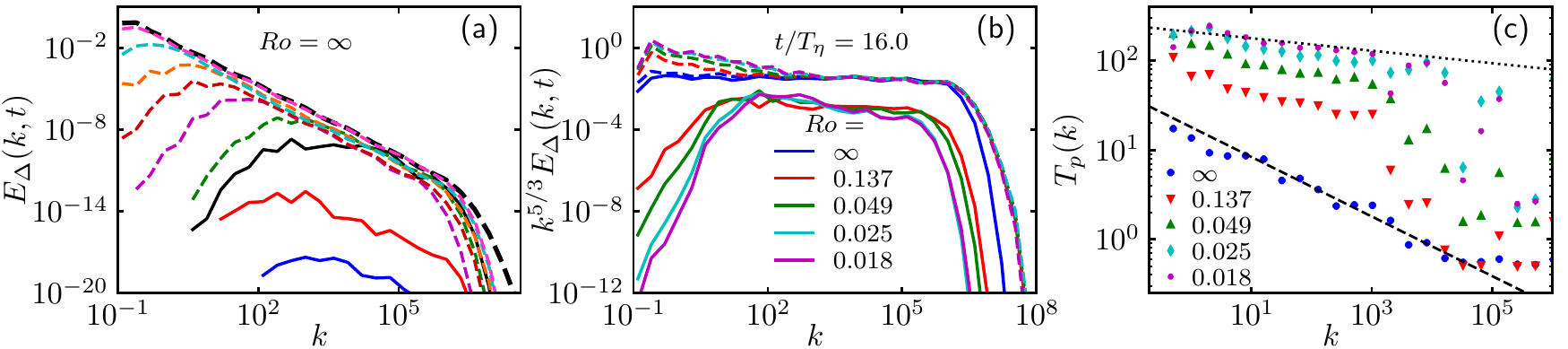}
	\caption{(a) Loglog plots of the ensemble averaged error spectra $E_\Delta(k, t)$ for $\mathrm{Ro} = \infty$ (i.e., no rotation) at different times $t/T_{\eta} = 0.23,  0.43,  0.5 ,  0.58,  0.99,  1.97,  3.94,  7.88$, and $15.76$ (from bottom to top). The solid and the dashed lines represent the error spectra in the exponential and the linear growth stages, respectively. The thick dashed black line represents the energy spectrum $E(k) \sim k^{-5/3}$. 
	(b) Loglog plots of the ensemble averaged error spectra compensated with $k^{5/3}$ for different $\mathrm{Ro}$ values (see legend) at a representative time $t/T_{\eta} = 16$. The dashed line with colors corresponding to the respective $\mathrm{Ro}$ represent the energy spectra $E(k) \sim k^{-2}$ ($k < k_\Omega$) and $E(k) \sim k^{-5/3}$ ($k > k_\Omega$), where $k_\Omega$ is the respective Zeman wavenumber. 
	(c)  Loglog plots of the predictability time $T_p(k)$ for different $\mathrm{Ro}$ (see legend). The black dashed line represents the scaling $T_p(k) \sim k^{-1/3}$ and the black dotted line scales as $T_p(k) \sim k^{-0.07}$. }
	\label{fig:error_spectra}
\end{figure*}

After investigating the predictability for the full system, we turn to investigate the scale-by-scale predictability of rotating turbulence. 
In Fig.~\ref{fig:error_spectra}(a), we plot the evolution of ensemble averaged error spectrum $E_\Delta(k, t)$ for $\mathrm{Ro} = \infty$ (non-rotating case). The solid lines represent the error spectra at different times during the exponential growth of the error. The error initially localized at the small scales $k_n \ge k_\eta$ grows exponentially to the orders of energy at the Kolmogorov scale in a short time $t \sim 0.5 T_{\eta}$, as shown by the error spectrum represented by the black solid line in Fig.~\ref{fig:error_spectra}(a). Then the error further grows to cascade towards the large scales---through the nonlinear interactions of the scales---as shown by the error spectra represented by the dashed lines at different times. The error spectrum saturates to the energy spectrum $E(k) \sim k^{-5/3}$ (shown by thick dashed black line) for the wavenumbers $k > k_L$ such that the reference and the perturbed trajectories completely decorrelate at the scales $ < k_L^{-1}$. Furthermore, we observe that the error spectrum at large scales is significantly affected and decorrelated in a relatively large amount of time due to the cascading of the error through nonlinear mechanisms: The true predictability time is much longer than the timescale of the exponential stage, rendering the Lyapunov exponent based predictability time irrelevant and insufficient for the scales in the inertial range. For example, the error spectrum at the large scales close to the forcing scales decorrelate in a large time $t \simeq 16 T_{\eta}$, making the dynamics at these large scale much more predictable.

Figure ~\ref{fig:error_spectra}(b) shows the ensemble averaged error spectra for various $\mathrm{Ro}$ values at the time $t = 16 T_\eta$ when the trajectories for $\mathrm{Ro} = \infty$ are completely decorrelated. We also plot, for reference, the energy spectra (represented by the respective colored dashed lines) $E(k) \sim k^{-2}$ ($k < k_\Omega$) and $E(k) \sim k^{-5/3}$ ($k > k_\Omega$), where $k_\Omega = \sqrt{\Omega^3/\varepsilon}$ is the respective Zeman wavenumber~\cite{Zeman1994} for the same $\mathrm{Ro}$ values. We observe that the error spectra for different finite $\mathrm{Ro}$ are of the same shape in the scales smaller than the Zeman scale ($k > k_\Omega$), though not decorrelated completely with the respective energy spectra. However, the error spectra in the Zeman range ($k < k_\Omega$) are different such that the time taken to saturate the error to the energy spectra increases with the decreasing $\mathrm{Ro}$. This also hints at the possible enhancement of the predictability due to the dynamics in the Zeman range. 

The time it takes for an error at a small scale (say, the Kolmogorov scale) to decorrelate the trajectories at a larger scale $L \sim k_L^{-1}$ is the predictability time associated with the scale $L$. We measure the predictability time at a given scale $l \sim k^{-1}$ by determining the time $T_p(k)$ such that the relative error spectrum $E_\Delta(k, T_p)/E(k) = \gamma$, where $\gamma \simeq 1$. Using the aforementioned definition, we compute an ensemble averaged predictability time $T_p(k)$ with $\gamma = 0.8$. In Fig.~\ref{fig:error_spectra}(c), The predictability timescales as $k^{-1/3}$ for $\mathrm{Ro} = \infty$, confirming the scaling reported in DNS study of homogeneous isotropic turbulence in an Eulerian approach~\cite{Berera2018}. This is, however, in contrast with the scaling $k^{-2/3}$ inferred from the Lorenz argument~\cite{Lorenz1969} and reported in DNS studies of 2D and 3D turbulence~\cite{Boffetta2001,Boffetta2017}.
For finite Ro, however, the predictability time tends to become scale independent, predominantly in the Zeman range, as Ro decreases. As we see in Fig.~\ref{fig:error_spectra}(c), the predictability time approaches a plateau, particularly in the Zeman range, as the Rossyb number is decreased.
This is evidenced by the close to \textit{zero} scaling exponent of the scaling $T_p(k) \sim k^{-0.07}$ for small $\mathrm{Ro} = 0.018$ [see the dotted line in Fig.~\ref{fig:error_spectra}(c)].
Furthermore, the predictability time $T_p(k)$, for small Ro, exhibits a transition from small to large predictability time around the Zeman scale.

\subsection{Finite Size Lyapunov Exponent}

The large-scale predictability is better described in terms of the \textit{finite size Lyapunov exponent} (FSLE) $\Lambda(\delta)$,  which is the average divergence rate of two trajectories initially separated by a \textit{finite} measure $\delta$ that is of the order of velocities in the inertial range~\cite{Aurell1996a,Aurell1996,Aurell1997,Cencini2013}. In particular, $\Lambda(\delta)$ is measured in terms of the (ensemble) average of the time $T_r(\delta)$ (the $r-$folding time) it takes for a separation $\delta \in \delta u(t) = \sqrt{2E_\Delta(t)}$ to grow by a factor $r$ to $r\delta$:
\begin{equation}
\Lambda(\delta) \equiv \frac{\ln r}{\langle T_r(\delta) \rangle},
\label{eqn:fsle}
\end{equation}
where $r$ is typically $O(1)$ $(r > 1)$, and the averaging $\langle \rangle$ is over an ensemble of many realizations. The FSLE was conceptualized to measure chaos and predictability in extended systems with many characteristic timescales and systems whose evolution involves nonlinear interactions among their degrees of freedom, such as turbulence~\cite{Aurell1996a,Boffetta2002a}. In the limit of $\delta \rightarrow 0$, the usual maximum Lyapunov exponent is recovered from the FSLE.

We compute $\Lambda(\delta)$ by fixing a sequence of separations $\delta_m = \delta_0 r^m$, where $m$ is an integer less than $M$ determined by $\delta_M = \delta u(t)$ at saturation, $r = \sqrt{2}$, and $\delta_0 \sim 10^{-4}$ is the initial separation to construct the sequence along $\delta u(t)$. We measure the time it takes to increase the separation from $\delta_m$ to $\delta_{m+1}$ and average it over the ensemble to obtain the FSLE $\Lambda(\delta_m)$, for each m, from the Eq.~\ref{eqn:fsle}. The predictability time from the initial separation $\delta_0$ to a given tolerance $\delta$ in uncertainty is then calculated as follows:
\begin{equation}
T_p(\delta_0, \delta)  = \int_{\delta_0}^{\delta} \frac{\mathrm{d} \ln \delta^\prime}{\Lambda(\delta^\prime)}.
\label{eqn:pt_fsle}
\end{equation}

\begin{figure}
	\includegraphics[scale=0.98]{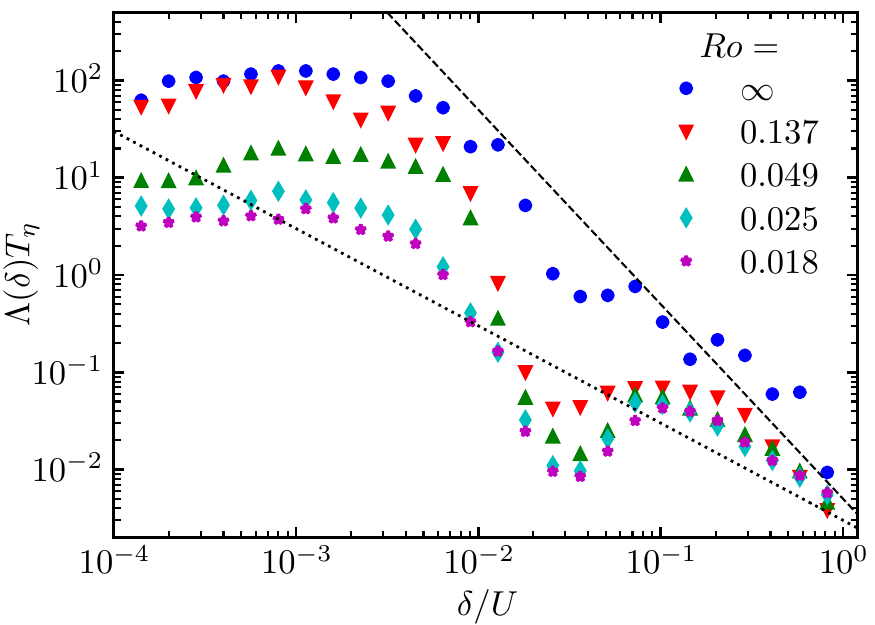}
	\caption{Loglog plots of the finite size Lyapunov exponent (FSLE) $\Lambda(\delta)$ obtained from the $r-$folding time averaged over $5000$ different realizations from our simulations of shell model. The black dashed line scales as $\delta^{-2}$ and the black dotted line as $\delta^{-1}$.}
	\label{fig:fsle_shellmodel}
\end{figure}

In Fig.~\ref{fig:fsle_shellmodel}, we plot the variation of the FSLE $\Lambda(\delta)$ with $\delta$. For $Ro = \infty$, we recover the standard scaling of the FSLE $\Lambda(\delta) \sim \delta^{-2}$, for $\delta$ in the inertial range of 3D turbulence\cite{Aurell1996a,Cencini2013}. For finite Rossby numbers, the $\Lambda(\delta)$ shows a dip for small $\delta$ in the inertial range, which corresponds to the newly developed frozen stage of error evolution in the rotating turbulence. 
This dip makes a significant contribution to the integral of the predictability time $T_p(\delta_0, \delta)$ in Eq.~\ref{eqn:pt_fsle} for $\delta$ in the inertial range, which results in a much longer predictability even for small $\delta$ in the inertial range, i.e., at small scales, as observed in the non-Zeman range of wavenumbers ($k > k_\Omega$) in Fig.~\ref{fig:error_spectra}(c). 

Furthermore, we find that the FSLE scales as $\delta^{-1}$ for large $\delta$ of the order of velocities in the Zeman range, for small values of Ro. From dimensional considerations and Lorenz argument, the FSLE $\Lambda(\delta) \sim k_L \norm{u(k_L)}$ for a separation $\delta$ of the order of $\norm{u(k_L)}$ in the Zeman range. Because $\norm{u(k)} = k^{-1/2}$ from the energy spectrum $E(k) \sim k^{-2}$ in the Zeman range, the FSLE $\Lambda(\delta) \sim \delta^{-1} $ and thus the scaling obtained from our shell model simulations is consistent with the aforementioned dimensional prediction.

It is worth noting that the predictability time for small $\delta$ increases as the Ro decreases, but this has very little effect on the large-scale predictability (see Fig.~\ref{fig:fsle_shellmodel}).

\section{Conclusion}
In conclusion, we studied the statistical properties of predictability in rotating turbulence by simulating two initially close realizations of velocity trajectories in the state space of a GOY shell model of rotating turbulence. Because of the chaotic nature of the flow, the trajectories separate at an exponential rate dependent on the Rossby number for an initial short time. For finite Ro, the error growth freezes for a time period determined by the Rossby number, and then the error grows algebraically until the trajectories completely decorrelate; this is peculiar to rotating turbulence. 
 
The large-scale predictability in rotating turbulence is found to increase with increasing rotation rate as predicted by the Lorenz argument that considers the growth of an uncertainty occurring on the nonlinear timescales. Interestingly, the predictability time for the full system is a power law in the Rossby number Ro. Furthermore, the predictability for large rotation rates tends to become scale independent, which contradicts the previously discussed prediction $T_L \sim k_L^{-1/2}$ based on the Lorenz argument. For finite Ro, the frozen stage of the error evolution is captured as a sudden dip in the FSLE plot, indicating that the time period of the frozen stage accounts for a considerable part of the predictability time. The dimensional predictions of $\Lambda(\delta) \sim \delta^{-1}$ for large $\delta$ in the Zeman range is validated for small Rossby numbers.

The frozen stage in the error evolution is perhaps the manifestation of the globally imposed timescale by the Coriolis force term modifying the character of the slow and fast dynamics in the state space of the shell model. The dip and the subsequent peak in the FSLE plot are signatures of the slow and the fast dynamics in the inertial range, respectively\cite{Mitchell2012}.
We speculate that the enhancement of predictability can be attributed to the coherent dynamics in the columnar structures formed in real rotating turbulence as predicted in theory (Taylor-Proudman theorem) and observed in direct numerical simulations~\cite{Sharma2018,Thiele2009,Godeferd2015}. 

The scale independence of predictability in the Zeman range for small Rossby numbers requires closer examination. This is because, for the increase and the scale independence of the predictability to happen simultaneously, the latter suggests that there is a nonlocal transfer of error among the large scales, although at a much slower rate. In the future, we want to investigate this property of the predictability in the Zeman range and the form of the algebraic growth of the error in detail. Finally, we think that our findings on the error dynamics and the predictability of rotating turbulence are of general interest in the study of extended dynamical systems with an imposed global timescale, such as stratified turbulence with a global timescale imposed by buoyancy.

\section*{Acknowledgment}
The author thanks to Sagar Chakraborty, Samriddhi Sankar Ray, and Manohar Kumar Sharma for valuable discussions and suggestions. The author would also like to thank Anando Gopal Chatterjee for his help in developing an alternative DNS code for this work. The simulations were performed on HPC2013 clusters of IIT Kanpur and using the resources provided by PARAM Sanganak under the National Supercomputing Mission, Government of India at IIT Kanpur.
	\bibliography{rathor_manuscript}
\end{document}